\documentclass[aps,superscriptaddress,twocolumn,nofootinbib,prd]{revtex4-1}
\usepackage{graphicx, epsfig, natbib, color, enumerate}

\newcommand{\alm}{a_{\ell m}}
\newcommand{\Ylm}{Y_{\ell m}}

\newcommand{\apjl}{ApJL}
\newcommand{\apjs}{ApJS}
\newcommand{\mnras}{MNRAS}
\newcommand{\physrev}{Phys. Rev.}
\newcommand{\physrevlett}{Phys. Rev. Lett.}

\newcommand{\etal}{\textit{et al.}}

\begin{document}

\title{Large-angle anomalies in the CMB}

\author{Craig J. Copi}
\affiliation{CERCA/Department of Physics, Case Western Reserve University, Cleveland, 
  OH 44106-7079}

\author{Dragan Huterer}
\affiliation{Department of Physics, University of Michigan, 
450 Church St, Ann Arbor, MI 48109-1040}

\author{Dominik J. Schwarz}
\affiliation{Fakult\"at f\"ur Physik, Universit\"at Bielefeld, Postfach 100131, 
  33501 Bielefeld, Germany}

\author{Glenn D. Starkman}
\affiliation{CERCA/ISO/Department of Physics, Case Western Reserve University, Cleveland, 
  OH 44106-7079}

\date{\today}

\begin{abstract}
  We review the recently found large-scale anomalies in the maps of
  temperature anisotropies in the cosmic microwave background. These include
  alignments of the largest modes of CMB anisotropy with each other and with
  geometry and direction of motion of the Solar System, and the unusually low
  power at these largest scales. We discuss these findings in relation to
  expectation from standard inflationary cosmology, their statistical
  significance, the tools to study them, and the various attempts to explain
  them.
\end{abstract}

\maketitle

\section{Introduction: Why large scales are interesting?}

The Copernican principle states that the Earth does not occupy a special place
in the universe and that observations made from Earth can be taken to be
broadly characteristic of what would be seen from any other point in the
universe at the same epoch.  The microwave sky is isotropic, apart from a
Doppler dipole and a microwave foreground from the Milky Way. Together with
the Copernican principle and some technical assumptions, an oft-inferred
consequence is the so-called cosmological principle.  It states that the
distributions of matter and light in the Universe are homogeneous and
isotropic at any epoch and thus also defines what we mean by cosmic time.

This set of assumptions is a crucial, implicit ingredient in obtaining most
important results in quantitative cosmology.  For example, it allows us to
treat cosmic microwave background (CMB) temperature fluctuations in different
directions on the sky as multiple probes of a single statistical ensemble,
leading to the precision determinations of cosmological parameters that we
have today.

Although we have some observational evidence that homogeneity and isotropy are
reasonably good approximations to reality, neither of these are
actual logical consequences of the Copernican principle. For example the
geometry of space could be homogeneous but anisotropic --- like the surface of
a sharp mountain ridge, with a gentle path ahead but the ground dropping
steeply away to the sides.  Indeed, three-dimensional space admits not just
the three well known homogeneous isotropic geometries (Euclidean, spherical and
hyperbolic -- $E^3$, $S^3$ and $H^3$),  but five others which are homogeneous
but anisotropic.  The two simplest are $S^2\times E^1$ and $H^2\times E^1$.
These spaces support the cosmological principle but have preferred directions.

Similarly, although the Earth might not occupy a privileged place in the
universe, it is not necessarily true that all points of observation are
equivalent.  For example, the topology of space may not be simply-connected
--- we could live in a three-dimensional generalization of a torus so that if
you travel far enough in certain directions you come back to where you
started.  While such three-spaces generically admit locally homogeneous and
isotropic geometries, certain directions or points might be singled out when
non-local measurements are considered.  For example the length of the shortest
closed non-trivial geodesic through a point depends on the location of that
point within the fundamental domain.  Similarly, the inhomogeneity and
anisotropy of eigenmodes of differential operators on such spaces are likely
to translate into statistically inhomogeneous and anisotropic large scale
structure, in the manner of Chladni figures on vibrating plates.

The existence of non-trivial cosmic topology and of anisotropic geometry are
questions that can only be answered observationally.  In this
regard, it is worth noting that our record at predicting the gross properties
of the universe on large scales from first principles has been rather poor.
According to the standard concordance model of cosmology, over $95$\% of the
energy content of the universe is extraordinary --- dark matter or dark energy
whose existence has been inferred from the failure of the Standard Model of
particle physics plus General Relativity to describe the behavior of
astrophysical systems larger than a stellar cluster --- while the very
homogeneity and isotropy (and inhomogeneity) of the universe owe to the
influence of an inflaton field whose particle-physics-identity is completely
mysterious even after three decades of theorizing.

The stakes are set even higher with the recent discovery of dark energy that
makes the universe undergo accelerated expansion. It is known that dark energy
can affect the largest scales of the universe --- for example, the clustering
scale of dark energy may be about the horizon size today. Similarly,
inflationary models can induce observable effects on the largest scales via
either explicit or spontaneous violations of statistical isotropy.
It is reasonable to suggest that statistical isotropy and homogeneity
should be substantiated observationally, not just assumed.  More generally,
testing the cosmological principle should be one of the key goals of modern
observational cosmology.

With the advent of high signal-to-noise maps of the cosmic microwave
background anisotropies and with the conduct of nearly-full-sky deep galaxy
surveys, statistical isotropy \textit{has} begun to be precisely tested.
Extraordinary full-sky temperature maps produced by the Wilkinson Microwave
Anisotropy Probe (WMAP), in particular, are revolutionizing our ability to
probe the universe on its largest scales \cite{Bennett2003,
  Spergel2003,Hinshaw2003,Spergel2006,wmap5,wmap7}.  In the near future, these
will be joined by higher resolution temperature maps and high-resolution
polarization maps and, eventually, by deep all-sky surveys, and perhaps by
tomographic 21-cm line observations that will extend our detailed knowledge of
the universe's background geometry and fluctuations into the interior of the
sphere of last scattering.

\begin{figure}
\includegraphics[width=\linewidth]{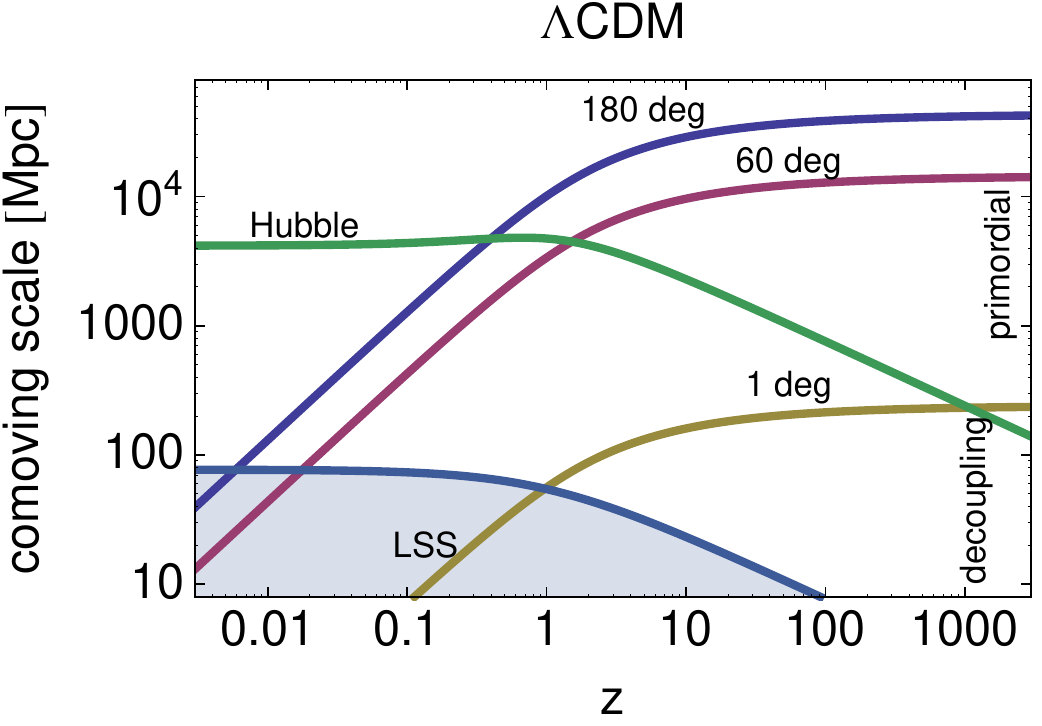}
\caption{The comoving length within the context of the concordance model of an
  arc seen at an fixed angle and the comoving Hubble length as functions of
  redshift. Linear perturbation theory is expected to work well outside the
  shaded region, in which the large scale structure (LSS)
  forms.} \label{fig:comscale}
\end{figure}

In this brief review, we describe the large-scale anomalies in the CMB data,
some of which were first reported on by the Cosmic Background Explorer (COBE)
Differential Microwave Radiometer (DMR) collaboration in the mid 1990's. In
particular, we report on alignments of the largest modes of CMB anisotropy
with each other, and with geometry and direction of motion of the Solar
System, as well as on unusually low angular correlations at the largest
angular scales. We discuss these findings and, as this is not meant to be a
comprehensive review and we emphasize results based on our own work
  in the area, we refer the reader to literature for 
all developments in the field. This review extends an earlier review on the
subject by \citet{Huterer_NewAst_review}, and complements another
review on statistical isotropy in this Special Issue \cite{Abramo_AdvAstro}.

The paper is organized as follows. In Sec.~\ref{sec:expect} we describe the
statistical quantities that describe the CMB, and the expectations for their
values in the currently favored $\Lambda$CDM cosmological model. In
Sec.~\ref{sec:align}, we describe the alignments at the largest scales, as
well as multipole vectors, which is a tool to study them. In
Sec.~\ref{sec:2pt}, we describe findings of low power at largest scales in the
CMB. Section \ref{sec:explain} categorizes and covers the variety of possible
explanations for these anomalies. We conclude in Sec.~\ref{sec:discussion}.

\section{Expectations from cosmological inflation} \label{sec:expect}

A fixed angular scale on the sky probes the physics of the universe at a range
of physical distances corresponding to the range of observable redshifts. This
is illustrated in Fig.~\ref{fig:comscale}, where the comoving lengths of arcs
at fixed angle are shown as a function of redshift, together with the comoving
Hubble scale. Angles of $1$ degree and less probe events that were in causal
contact at all epochs between the redshift of decoupling and today; this
redshift range includes physical processes such as the secondary CMB
anisotropies. The situation is different for angles $> 60$ degrees, which
subtend arcs that enter our Hubble patch only at $z \lesssim 1$. Therefore,
the primordial CMB signal on such large angular scales could only be modified
by the physics of local foregrounds and cosmology in the relatively recent
past ($z\lesssim 1$).  Because they correspond to such large physical scales,
the largest observable angular scales provide the most direct probe of the
primordial fluctuations --- whether generated during the epoch of cosmological
inflation, or preceding it.

\subsection{Statistical isotropy}

What do we expect for the large angular scales of the CMB? A crucial
ingredient of cosmology's concordance model is cosmological inflation --- a
period of accelerating cosmic expansion in the early universe.  If we assume
that inflationary expansion persisted for sufficiently many e-folds, then we
expect to live in a homogeneous and isotropic universe within a domain larger
than our Hubble volume. This homogeneity and isotropy will not be exact, but
should characterize both the background and the statistical distributions of
matter and metric fluctuations around that background.  These fluctuations are
made visible as anisotropies of the CMB temperature and polarization, which
are expected to inherit the underlying statistical isotropy. The temperature
$T$ seen in direction $\hat e$ is predicted to be described by a Gaussian
random field on the sky (i.e.\ the 2-sphere $S^2$), which implies that we can
expand it in terms of spherical harmonics $Y_{\ell m}(\hat e)$ multiplied by
independent Gaussian random coefficients $a_{\ell m}$ of zero mean.
  
Statistical isotropy implies that the expectation values of all $n$-point
correlation functions (of the temperature or polarization) are invariant under
arbitrary rotations of the sky. As a consequence the expectation of the
temperature coefficients is zero, $\langle a_{\ell m}\rangle =0$, for all
$\ell > 0$ and $m = -\ell, -\ell + 1, \dots, + \ell$. The two-point
correlation becomes a function of $\cos \theta \equiv \hat e_1 \cdot \hat e_2$
only, and can be expanded in terms of Legendre polynomials
\begin{equation}
  \langle T(\hat e_1) T(\hat e_2)\rangle \equiv 
  C(\theta) = \frac{1}{4\pi}\sum_\ell (2\ell +1) C_\ell P_\ell (\cos \theta).
  \label{eqn:PlCtheta}
\end{equation}
Statistical independence implies that expectations 
of $a_{\ell m}$ with different $\ell$ and $m$ vanish.  In
particular, the two-point correlation function is diagonal in $\ell$ and $m$
\begin{equation}
\langle a^*_{\ell m} a_{\ell' m'} \rangle \propto \delta_{\ell \ell'} \delta_{m m'} .
\end{equation}
Statistical isotropy adds that the constant of proportionality depends only on
$\ell$, not $m$
\begin{equation}
\langle a^*_{\ell m} a_{\ell' m'} \rangle = \delta_{\ell \ell'}
\delta_{m m'} C_\ell.
\end{equation}
The variance $C_\ell$ is called the angular power of the multipole $\ell$.
The higher $n$-point correlation functions are constrained in similar ways
but, as we will see below, are not expected to provide independent
information if a simple inflationary scenario was realized by nature.

\subsection{Gaussianity}

If inflation is driven by a single dynamically relevant degree of freedom with
appropriate properties (minimal coupling, Minkowski vacuum in UV limit, etc.),
then we can reduce the quantization of matter and space-time fluctuations
during inflation to the problem of quantizing free scalar fields. For free
fields the only non-trivial object is the two-point correlation (the
propagator), and all higher correlation functions either vanish or are just
some trivial combination of the two-point function.  This property is mapped
onto the temperature field of the CMB. A classical random field with these
properties is a Gaussian with mean $T_0$ and variance $C(\theta)$. Thus the
brightness of the primordial CMB sky is completely characterized by $T_0$ and
$C(\theta)$ (or $C_\ell$).  Note that evolution of perturbations leads to
deviations from Gaussianity that would mostly be evident at very small scales
($\ell \gg 100$).  Moreover, many inflationary models predict small deviations
from Gaussianity; these are described in other contributions to this volume
\cite{Byrnes_AdvAstro,Chen_AdvAstro}.

\subsection{Scale-invariance} 

Another generic feature of inflation is the almost scale-invariance of the
power spectrum of fluctuations.  This can be understood easily, as the Hubble
scale is approximately constant during inflation as the wavelengths of
observable modes are redshifted beyond the horizon. Given that fluctuations of
modes on horizon exit are related to the Hubble parameter, $\delta \phi =
H/2\pi$, these modes have similar amplitudes. However, scale-invariance is not
exact. In canonical slow-roll inflation models, the deviation from exact
scale-invariance is due to the evolution of the Hubble parameter during
inflation, which is measured by the so-called first slow-roll function
$\epsilon_1 \equiv \dot{d}_{\mathrm{H}}$ where $d_{\mathrm{H}}\equiv H^{-1}$
is the Hubble distance.  From the weak energy condition $\epsilon_1> 0$, while
$\epsilon_1 \ll 1$ during slow roll inflation.

At the level of the angular power spectrum, exact scale-invariance implies the
Sachs-Wolfe ``plateau'' (i.e.\ constancy of $l(l+1)C_\ell$ at low $\ell$)
\cite{Sachs:1967er}
\begin{equation}
C_\ell = {2\pi A\over \ell (\ell +1)}.
\end{equation}
Here, again in the slow-roll parametrization, $A \sim
(H_{\mathrm{infl}}/M_{\mathrm{P}})^2 T_0^2/\epsilon_1$. This neglects
secondary anisotropies like the late
time, integrated Sachs-Wolfe effect (particularly important at very low
$\ell$) and
the contribution from gravitational waves. 
Furthermore, inflation predicts a small departure from scale invariance, which
has recently been detected (e.g.\ \cite{wmap7}), and which also contributes to
a tilt in the aforementioned plateau.

\begin{figure}
\includegraphics[width=\linewidth]{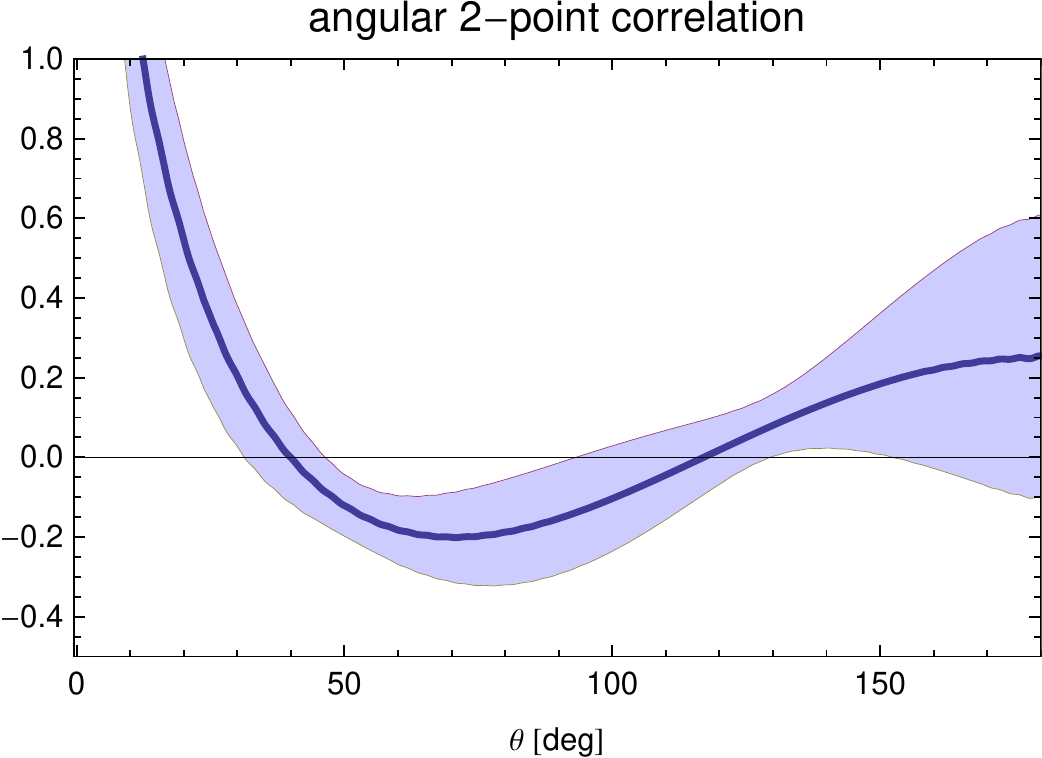}
\caption{Mean and (cosmic) variance of the angular two-point correlation
  function as expected from cosmological inflation (arbitrary normalization).
  Only statistical isotropy, Gaussianity and scale-invariance are assumed.
  Tensors, spectral tilt, reionization and the integrated Sachs-Wolfe effect
  are neglected for the purpose of this plot.  Comparison to the prediction
  from the best-fit $\Lambda$CDM model (Fig.~\ref{fig:ctheta}) reveals that
  these corrections are subdominant.  Note that cosmic variance errors at
  different values of $\theta$  are  very highly correlated.} \label{fig:2point}  
\end{figure}

\subsection{Cosmic variance}

As we can measure only one sky, it is important to find the best estimators of
$C_\ell$ and $C(\theta)$. Let us for the moment assume that we are able to
measure the primordial CMB of the full sky, without any instrumental noise.
We also restrict ourselves to $\ell \geq 2$, as the variance of the monopole
cannot be defined and the measured dipole is dominated by our motion through
the universe, rather than by primordial physics.  (Separation of the Doppler
dipole from the intrinsic dipole is possible in principle
\cite{Peebles:1968zz,Kamionkowski:2002nd}, but not with existing data.)
Statistical isotropy suggests to estimate the angular power by
\begin{equation}
\hat{C}_\ell = \frac{1}{2\ell +1} \sum_{m = -\ell}^{+\ell} |a_{\ell m}|^2,
\end{equation} 
which satisfies $\langle \hat{C}_\ell \rangle = C_\ell$ and is thus
unbiased. The variance of this estimator can be calculated assuming
Gaussianity:
\begin{equation}
\mathrm{Var}(\hat{C}_\ell) = \frac{2}{2\ell + 1} \hat{C}_\ell^2.
\end{equation} 
It can be shown that, assuming statistical isotropy and Gaussianity, 
$\hat{C}_\ell$ is the best estimator in the sense that it has minimal variance
and is unbiased.  However, we emphasize that these qualities depend
intrinsically on the correctness of the underlying assumptions.

With these same assumptions, the variance of the two-point correlation
function is easily shown to be
\begin{equation}
\mathrm{Var}[\hat{C}(\theta)] = \frac{1}{8\pi^2} \sum_\ell (2\ell + 1) C_\ell^2 P_\ell^2(\cos \theta),
\end{equation} 
where $\hat{C}(\theta)$ is calculated from $\hat{C}_\ell$ following
Eq.~(\ref{eqn:PlCtheta}).

Putting the results of this section together allows us to come up with a
generic prediction of inflationary cosmology for $C(\theta)$ on the largest
angular scales; see Fig.~\ref{fig:2point}.

\section{Alignments}
\label{sec:align}

In brief, the upshot of the previous section is that the twin assumptions of
statistical isotropy and Gaussianity are the starting point of \textit{any}
CMB analysis. The measurements of the CMB monopole, dipole and $(\Delta
T)_{\mathrm{rms}}$ tell us that isotropy is observationally established at the
per cent level without any cosmological assumption, and at a level $10^{-4}$
if we attribute the dominant contribution to the dipole to our peculiar
motion.  For the purpose of cosmological parameter estimation, the task is to
test the statistical isotropy of the CMB brightness fluctuations.  At the
largest angular scales, this can only be done by means of full sky maps.

Let us assume that the various methods that have been developed to get rid of
the Galactic foreground in single frequency band maps of the microwave sky are
reliable (though we argue below that this might not be the case).  Our review
of alignments will be based on the internal linear combination (ILC) map
produced by the WMAP team, which is based on a minimal variance combination of
the WMAP frequency bands.  The weights for the five frequency band maps are
adjusted in 12 regions of the sky, one region lying outside the Milky Way and
11 regions along the Galactic plane.

\subsection{Multipole vectors}
\label{subsec:mv}

\begin{figure}
\includegraphics[width=\linewidth]{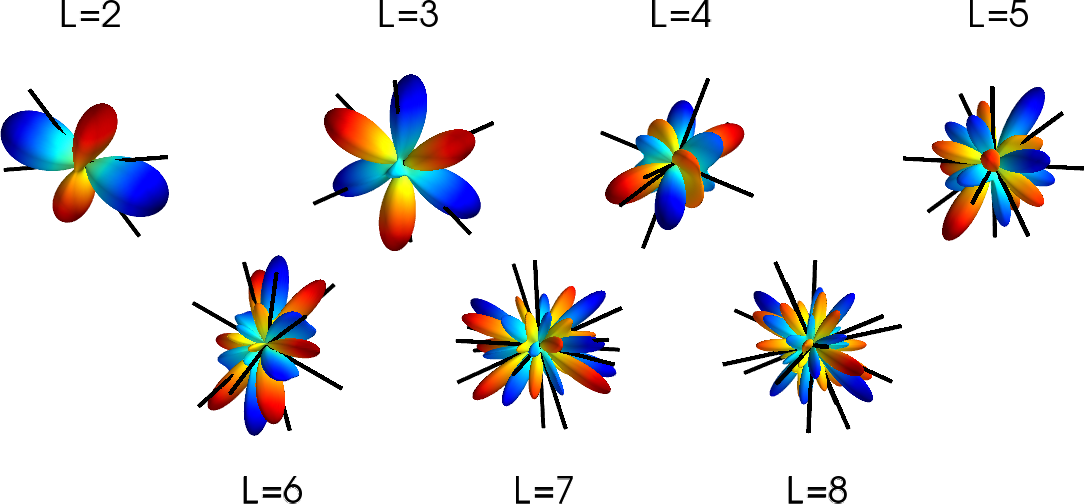}
\caption{Multipole vectors of our sky, based on WMAP five year
  full-sky ILC map and with galactic plane coinciding with the plane of the
  page. The temperature pattern at each multipole $\ell$ ($2\leq \ell\leq 8$)
  can either be described by an angular temperature pattern (colored lobes in
  this figure), or alternatively by precisely $\ell$ multipole vectors
  (black ``sticks''). While the multipole vectors contain all
  information about the directionality of the CMB temperature pattern, they
  are not simply related to the hot and cold spots and, for
  example, do not correspond to the temperature minima/maxima
  \cite{lowl2}. Notice that $\ell=2$ and $3$ temperature patterns are rather
  planar with the same plane, and that their vectors lie approximately in this
  plane. Adopted from Ref.~\cite{Copi2004}. }
\label{fig:MV}
\end{figure}

To study the orientation and alignment of CMB multipoles, Copi, Huterer \&
Starkman~\cite{Copi2004} introduced to cosmology the use of multipole vectors;
an alternative representation of data on the sphere.  The multipole vectors
contain information about the ``directions'' associated with each multipole.
In this new basis the temperature fluctuation multipole, $\ell$, may be
expanded as
\begin{equation}
  T_\ell \equiv \sum_{m=-\ell}^\ell \alm \Ylm \equiv A^{(\ell)}\left[
    \prod_{i=1}^\ell (\hat v^{(\ell,i)}\cdot \hat e) - \mathcal{T}_\ell
    \right],
\label{eq:MV}
\end{equation}
where $\hat v^{(\ell,i)}$ is the $i^{\mathrm{th}}$ vector for the
$\ell^{\mathrm{th}}$ multipole, $\hat e$ is the usual radial unit vector,
$\mathcal{T}_\ell$ is the trace of the preceding product of multipole vector
terms, and $A^{(\ell)}$ is the ``power'' in the multipole.  By construction, we
immediately see that $T_\ell$ is a symmetric, traceless, rank $\ell$ tensor.
Subtracting $\mathcal{T}_\ell$ ensures that $T_\ell$ is traceless and the dot
products explicitly show this is a rotationally invariant quantity (a scalar
under rotations).  This form makes the symmetry properties obvious.  As an
example the quadrupole is written as
\begin{equation}
  T_2 = A^{(2)} \left[ (\hat v^{(2,1)}\cdot\hat e)(\hat v^{(2,2)}\cdot\hat
    e) - \frac13 \hat v^{(2,1)}\cdot\hat v^{(2,2)} \right].
\label{eq:trace}
\end{equation}

These two forms for representing $T_\ell$, harmonic and multipole-vector, both
contain the same information. At the same time, they are fundamentally
different from each other.  Each unit vector $\hat v^{(\ell,i)}$ has two
degrees of freedom while the scalar, $A^{(\ell)}$ has one; thus the multipole
vector representation contains the full $2\ell+1$ degrees of freedom.  Note
that we call each $\hat v^{(\ell,i)}$ a multipole vector but it is only
defined up to a sign.  We can always reflect the vector through the origin by
absorbing a negative sign into the scalar $A^{(\ell)}$; thus these vectors
actually are headless.  Regardless, we will continue to refer to them as
multipole vectors and not use the overall sign of the vector in our analysis.
This issue is equivalent to choosing a phase convention, such as the
Condon-Shortley phase for the spherical harmonics.  For the work reviewed
here the overall phase is not relevant and thus will not be specified.

An efficient algorithm to compute the multipole vectors for low-$\ell$ has
been presented in \cite{Copi2004} and is publicly available \cite{MV_code};
other algorithms have been proposed as well
\cite{Katz2004,Weeks04,Helling}. Interestingly, after the publication of
the CHS paper \cite{Copi2004}, Weeks \cite{Weeks04} pointed out that
multipole vectors were actually first used by Maxwell \cite{Maxwell}
more than 100 years ago in his study of multipole moments in
electrodynamics.   They remain in use in geometrology, nuclear physics,
and other fields.

The relation between multipole vectors and the usual harmonic basis is very
much the same as that between Cartesian and spherical coordinates of
standard geometry: both are complete bases, but specific problems are much
more easily addressed in one basis than the other. In particular, we and
others have found that multipole vectors are particularly well suited for
tests of planarity and alignment of the CMB anisotropy pattern. Moreover, a
number of interesting theoretical results have been found; for example,
Dennis \cite{Dennis2005} analytically computed the two-point correlation
function of multipole vectors for a Gaussian random, isotropic underlying
field.  Numerous quantities have been proposed for assigning directions to
multipoles and statistics on these quantities have been studied.  In Copi
\etal~\cite{lowl2} we have summarized these attempts and have shown their
connections to the multipole vectors.

\subsection{Planarity and Alignments}

Tegmark \etal~\cite{TOH} and de Oliveira-Costa \etal~\cite{deOliveira2004} first 
argued that the octopole is planar and that the quadrupole and octopole planes 
are aligned. In Schwarz \etal~\cite{Schwarz2004}, followed up by 
Copi \etal~\cite{lowl2,wmap123}, we investigated the quadrupole-octopole
shape and orientation using the multipole vectors.  The quadrupole is
fully described by two multipole vectors, which define a plane.   
This plane can be described by the ``oriented area'' vector
\begin{equation}
  \vec w^{(\ell;i,j)}\equiv \hat v^{(\ell, i)}\times \hat v^{(\ell, j)}.
\end{equation}
(Note that the oriented area vector \textit{does not} fully characterize the
quadrupole, as pairs of quadrupole multipole vectors related by a rotation
about the oriented area vector lead to the same oriented area vector.)  The
octopole is defined by three multipole vectors which determine (but again are
not fully determined by) three area vectors. Hence there are a total of four
planes determined by the quadrupole and octopole.

\begin{figure}
  \includegraphics[width=0.5\textwidth]{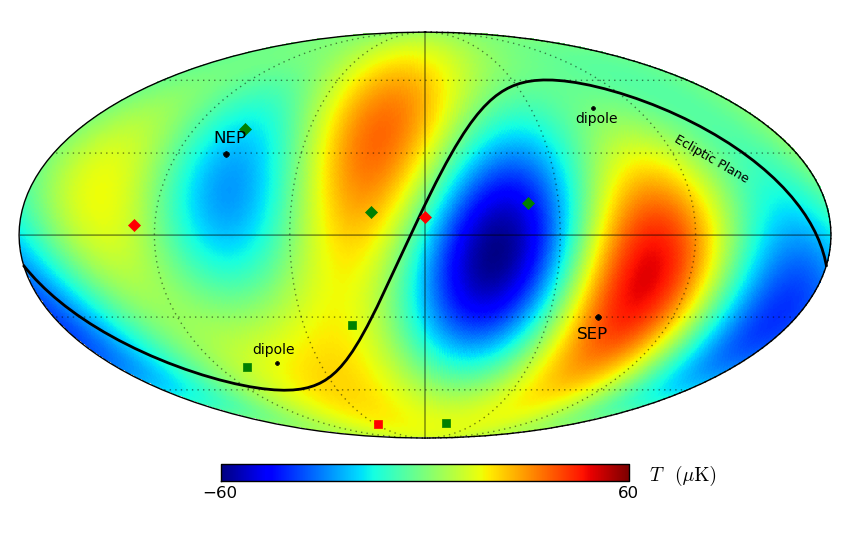}
  \caption{
    Quadrupole and octopole ($\ell=2$ and $3$) temperature
    anisotropy of the WMAP sky map in galactic coordinates, shown with the
    ecliptic plane and the cosmological dipole.  Included are the multipole
    vectors (solid diamonds); two for the quadrupole (red diamonds) and
    three for the octopole (green diamonds).  We also show the four
    normals (solid squares) to the planes defined by vectors that describe
    the quadrupole and octopole temperature anisotropy; one normal is
    defined by the quadrupole (red square) and three by the octopole
    (green squares). Note that three out of four normals lie very close
    to the dipole direction. The probability of this alignment being
    accidental is about one part in a thousand. Moreover, the ecliptic
    plane traces out a locus of zero of the combined quadrupole and
    octopole over a broad swath of the sky --- neatly separating a hot
    spot in the northern sky from a cold spot in the south. These apparent
    correlations with the solar system geometry are puzzling and currently
    unexplained.  }
\label{fig:align}
\end{figure}

In Copi \etal~\cite{wmap123} we found that (see Fig.~\ref{fig:align})
\begin{itemize}
\item the four area vectors of the quadrupole and octopole are mutually close
(i.e.\ the quadrupole and octopole planes are aligned) at the $99.6$\% C.L.;
\item the quadrupole and octopole planes are orthogonal to the ecliptic at the
  $95.9$\% C.L.; this alignment was at $98.5$\% C.L. in our analysis of the
  WMAP 1 year maps. The reduction of alignment was due to WMAP's adaption of a
  new radiometer gain model for the 3 year data analysis, that took seasonal
  variations of the receiver box temperature into account --- a systematic
  that is indeed correlated with the ecliptic plane. We regard that as clear
  evidence that multipole vectors are a sensitive probe of alignments;
  \item the normals to these four planes are aligned with the direction of the
cosmological dipole (and with the equinoxes) at a level inconsistent with
Gaussian random, statistically isotropic skies at $99.7$\% C.L.;
\item the ecliptic threads between a hot and a cold spot of the combined
  quadrupole and octopole map, following a node line across about $1/3$ of the
  sky and separating the three strong extrema from the three weak extrema of
  the map; this is unlikely at about the $95$\% C.L.
\end{itemize}
These numbers refer to the WMAP ILC map from three years of data; other maps
give similar results.
Moreover, correction for the kinematic quadrupole -- slight
  modification of the quadrupole due to our motion through the CMB rest frame
  -- must be made and increases significance of the alignments. See Table 3 of
  Ref.~\cite{wmap123} for the illustration of both of these points.

While not all of these alignments are statistically independent, their
combined statistical significance is certainly greater than their individual
significances.  For example, given their mutual alignments, the conditional
probability of the four normals lying so close to the ecliptic, is less than
2\%; the combined probability of the four normals being both so aligned with
each other and so close to the ecliptic is less than $0.4\% \times 2\% =
0.008\%$.  These are therefore clearly surprising, highly statistically
significant anomalies --- unexpected in the standard inflationary theory and
the accepted cosmological model.

Particularly puzzling are the alignments with solar system features.
CMB anisotropy should clearly not be correlated with our local
habitat. While the observed correlations seem to hint that there is
contamination by a foreground or perhaps by the scanning strategy of the
telescope, closer inspection reveals that there is no obvious way to
explain the observed correlations.   Moreover, if their explanation is
that they are a foreground, then that will likely exacerbate other
anomalies that we will discuss in section \ref{sec:missing} below.

Our studies (see \cite{lowl2}) indicate that the observed alignments are
with the ecliptic plane, with the equinox or with the CMB dipole, and
\textit{not} with the Galactic plane: the alignments of the
quadrupole and octopole planes with the equinox/ecliptic/dipole directions
are much more significant than those for the Galactic plane.  Moreover, it is
remarkably curious that it is precisely the ecliptic alignment that has
been found on somewhat smaller scales using the power spectrum analyses of
statistical isotropy \cite{Eriksen_asym,Hansen_asym,Bernui06,Bernui07}.

Finally, it is important to make sure that the results are unbiased by 
unfairly chosen statistics. We have studied this issue extensively in
\cite{lowl2}, and here we briefly describe the principal statistics used to
quantify the probability of alignments quoted just above.

To define statistics we first compute the three dot-products between the
quadrupole area vector and the three octopole area vectors,
\begin{equation}
  A_k \equiv |\vec w^{(2;1,2)} \cdot \vec w^{(3;i,j)}|.
\end{equation}
The absolute value is included since the multipole vectors are
headless; thus each $A_k$ lies in the interval $[0,1]$.
Two natural choices of statistics that are independent of the ordering of
$A_k$ are
\begin{eqnarray}
  S &\equiv& {1\over 3} \left (A_1 + A_2 + A_3\right ), \quad \mbox{and} \\
  T &\equiv& 1-{1\over 3}\left [(1-A_1)^2 + (1-A_2)^2 + (1-A_3)^2\right ].
  \nonumber
\end{eqnarray}
Both $S$ and $T$ statistics can be viewed as the suitably defined ``distance''
to the vertex $(A_1, A_2, A_3)=(0, 0, 0)$.  A third obvious choice,
$(A_1^2+A_2^2+A_3^2)/3$, is just $2S-T$. To test alignment of the quadrupole
and octopole planes with one another we quoted the $S$ statistic
numbers; $T$ gives similar results.

Alternatively, generalizing the definition in \cite{TOH}, one
can find, for each $\ell$, the choice, $n_\ell$, of $z$ axis that maximizes 
the angular momentum dispersion 
\begin{equation}
  \hat L^2_\ell \equiv { \sum_{m=-\ell}^\ell m^2 \vert a_{\ell m}\vert^2
    \over \ell^2 \sum_{m=-\ell}^\ell \left| \alm \right|^2 }.
  \label{eqn:angmomn}
\end{equation}
One can then compare the maximized value 
with that from simulated isotropic skies \cite{lowl2}.  
Because $\ell=2,3$ are both planar (the quadrupole trivially so, the
octopole because the three planes of the octopole are nearly parallel),
the direction that maximizes the angular momentum dispersion of 
each is nearly the same as the (average) direction of that multipole's
planes.  Thus, the alignment of the octopole and quadrupole can 
be seen either from the $\mathcal{S}$ statistic, or by looking
at the alignment of $n_2$ with $n_3$.

To test alignments of multipole planes with physical directions, we find the plane 
whose normal, $\hat{n}$, has the largest dot product with the sum of the four quadrupole
and octopole area vectors \cite{lowl2}. Again, since $\vec{w}_i \cdot \hat{n}$ is defined
only up to a sign, we take the absolute value of each dot product. Therefore,
we find the direction $\hat{n}$ that maximizes
\begin{equation}
  \mathcal{S}\equiv \frac 1 {N_{\ell}} \sum_{i=1}^{N_\ell} \left |\vec{w}_i
  \cdot \hat{n}\right |.
  \label{eq:S_single_mult}
\end{equation}

\subsection{Summary}

The study of alignments in the low-$\ell$ CMB has found a number of
peculiarities.  We have shown that the alignment of the quadrupole and
octopole planes is inconsistent with Gaussian, statistically isotropic
skies at least at the $99\%$ confidence level.  Further a number of (possibly
related) alignments occur at $95\%$ confidence levels or greater.  Put
together these provide a strong indication that the full sky CMB WMAP maps
are inconsistent with the standard cosmological model at large angles.
Even more peculiar is the alignment of the quadrupole and octopole with
solar system features (the ecliptic plane and the dipole).  

This is strongly suggestive of an unknown systematic in the data reduction;
however, careful scrutiny has revealed no such systematic (except the
mentioned modification of the radiometer gain model, that lead to a reduction
of ecliptic alignment); see Secs.~\ref{sec:analysis} and \ref{sec:instr} for
  further discussion of the data analysis and instrumental explanations.  We again
  stress that these results hold for full sky maps; maps that are produced
  through combination of the individual frequency maps in such a way as to
  remove foregrounds.  An alternative approach that removes the need for full
  sky maps is presented in the next section.

\section{two-point Angular Correlation Function}\label{sec:2pt}

The usual CMB analysis solely involves the spherical harmonic decomposition
and the two-point angular power spectrum.  There are many reasons for
this.  Firstly, when working with a statistically
isotropic universe the angular power spectrum contains all of the physical
information.  Secondly, the standard theory predicts the $\alm$ and their
statistical properties, through the $C_\ell$, thus the spherical harmonic
basis is a natural one to employ.  Finally, as measured today the angular
size of the horizon at the time of last scattering is approximately $1$
degree.  Since $\theta(\mathrm{deg})\simeq 200/\ell$ the causal physics at
the surface of last scattering leaves its imprint on the CMB on small
scales, $\theta\lesssim1^\circ$ or $\ell \gtrsim100$.  The two-point angular
power spectrum focuses on these small scales, making it a good means of
exploring the physics of the last scattering surface.  The tremendous
success of the standard model of cosmology has been the agreement of the
theory and observations on these small scales, allowing for the precise
determination of the fundamental cosmological parameters~\cite{WMAP5params}.

The two-point angular correlation function provides another means of analyzing
CMB observations and should not be ignored even if, in principle, it contains
the same information as the angular power spectrum.  Thus, even in the case of
full sky observations and/or statistical isotropy there are benefits in
looking at the data in different ways.  The situation is similar to a
function in one dimension where it is widely appreciated that features easily
found in the real space analysis can be very difficult to find in the Fourier
transform, and vice versa.  Furthermore, the two-point angular correlation
function highlights behavior at large angles (small $\ell$); the opposite of
the two-point angular power spectrum.  Thus the angular correlation function
allows for easier study of the temperature fluctuation modes that are
super-horizon sized at the time of last scattering.  Finally, the angular
correlation function in its simplest form is a direct pixel based measure (see
below).  Thus it does not rely on the reconstruction of contaminated regions
of the sky to employ.  This makes it a simple, robust measure even for partial
sky coverage.

\subsection{Definition}

Care should be taken when discussing statistical quantities of the CMB and
their estimators.  It rarely is in the literature.  Here we follow the
notation of Copi \etal~\cite{wmap123}, also see \cite{wmap12345}. The two
point angular correlation function, 
\begin{equation}
  \tilde{\mathcal{C}} (\hat e_1, \hat e_2) \equiv \langle T(\hat e_1)
  T(\hat e_2) \rangle,
  \label{eqn:Ctheta}
\end{equation}
is the \textit{ensemble average} (represented by the angle brackets,
$\langle\cdot\rangle$) of the product of the temperatures in the directions
$\hat e_1$ and $\hat e_2$.  Unfortunately we only have one universe to
observe so this ensemble average cannot be calculated.  Instead we average
over the sky so that what we mean by the two-point angular correlation
function is a \textit{sky average},
\begin{equation}
  \mathcal{C}(\theta) \equiv \overline{T(\hat e_1) T(\hat e_2)},
\end{equation}
where the average is over all pairs of pixels with $\hat e_1\cdot\hat
e_2=\cos\theta$.  This is a pixel based quantity and can be calculated for any
region of the sky (of course not all separations $\theta$ may be represented
on a given patch of the sky, depending on its geometry).

\subsection{Missing angular power at large scales}
\label{sec:missing}

Spergel \etal~\cite{Spergel2003}  found that the two-point correlation
function nearly vanishes on scales greater than about 60 degrees, contrary
to what the standard $\Lambda$CDM theory predicts, and in agreement with
the same finding obtained from COBE data about a decade earlier
\cite{DMR4}.

\begin{figure}[t]
  \includegraphics[scale=0.30]{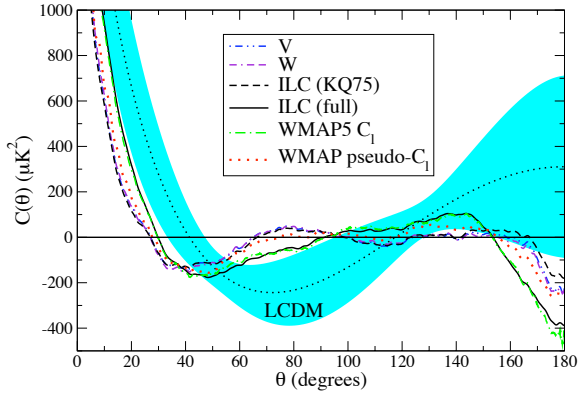}
  \caption{Two-point angular correlation function, ${\mathcal C}(\theta)$,
    computed in pixel space, for three different bands masked with the KQ75
    mask (from WMAP 5 year data). Also shown is the correlation function for
    the ILC map with and without the mask, and the value expected for a
    statistically isotropic sky with best-fit $\Lambda$CDM cosmology together
    with 68\% cosmic variance error bars.  Even by eye, it is apparent that
    masked maps have $C(\theta)$ that is consistent with zero at
    $\theta\gtrsim 60$ deg.  We also show the $C(\theta)$ computed from the
    ``official'' published maximum likelihood estimator-based
    $C_\ell$. Clearly, the MLE-based $C_\ell$, as well as ${\mathcal
      C}(\theta)$ computed from the full-sky ILC maps, are in significant
    disagreement with the angular correlation function computed from cut-sky
    maps. Adopted from Ref.~\cite{wmap12345}.}
\label{fig:ctheta}
\end{figure}

We have revisited the angular two-point function in the 3-yr WMAP data in
Ref.~\cite{wmap123} and the 5-yr WMAP data in \cite{wmap12345}; see
Fig.~\ref{fig:ctheta}.  From this figure we qualitatively see the
following:
\begin{itemize}
\item All of the cut-sky map curves are very similar to each other, and they
  are also very similar to the Legendre transform of the pseudo-$C_\ell$
  estimate of the angular power spectrum, which is not surprising given that
  the two are formally equivalent \cite{Pontzen_Peiris}.  Meanwhile the
  full-sky ILC $\mathcal{C}(\theta)$ and the Legendre transform of the maximum
  likelihood estimator (MLE) of the $C_\ell$ agree well with each other, but
  not with any of the others.
\item The most striking feature of the cut-sky (and pseudo-$C_\ell$)
  $\mathcal{C}(\theta)$, is that all of them are very nearly zero above
  about $60^\circ$, except for some anti-correlation near $180^\circ$. This
  is also true for the full-sky curves, but less so.
\end{itemize}

In order to be more quantitative about these observations we adopt the
$S_{1/2}$ statistic introduced by the WMAP team~\cite{Spergel2003} which
quantifies the deviation of the two-point correlation function from zero,
\begin{equation}
  S_{1/2} \equiv \int_{-1}^{1/2} \left[ \mathcal{C}(\theta)\right]^2 d
  (\cos\theta).
  \label{eqn:Shalf}
\end{equation}
Spergel \etal~\cite{Spergel2003} found that only $0.15\%$ of the parameter
sets in their Markov chain of $\Lambda$CDM model CMB skies had lower values
of $S_{1/2}$ than the observed one-year WMAP sky.

Applying this statistic we have found that the two-point function computed
from the various cut-sky maps shows an even stronger lack of power, for
WMAP 5 year maps significant at the $0.037$\%-$0.025$\% level depending on
the map used; see Fig.~(\ref{fig:ctheta}). However, we also found that,
while $\mathcal{C}(\theta)$ computed in pixel space over the masked sky
agrees with the harmonic space calculation that uses the pseudo-$C_\ell$
estimator, it disagrees with the $C_\ell$ obtained using the MLE (advocated
in the 3rd year WMAP release \cite{Spergel2006}). The MLE-based $C_\ell$
lead to $C(\theta)$ that is low (according to the $S_{1/2}$ statistic) only
at the $4.6$\% level.

There are actually two interesting questions one can ask:
\begin{enumerate}[(i)]
\item Is the correlation function measured on the cut sky compatible with cut-sky
expectation from the Gaussian random, isotropic underlying model?
\item Is the reconstruction of the full sky correlation function from partial
  information compatible with the expectation  from the Gaussian random,
  isotropic underlying model?
\end{enumerate}
Our results refer to the first question above. The second question, while also
extremely interesting, is more difficult to be robustly resolved because the
reconstruction uses assumptions about statistical isotropy (see the next
subsection).

\begin{figure}
  \includegraphics[width=0.5\textwidth]{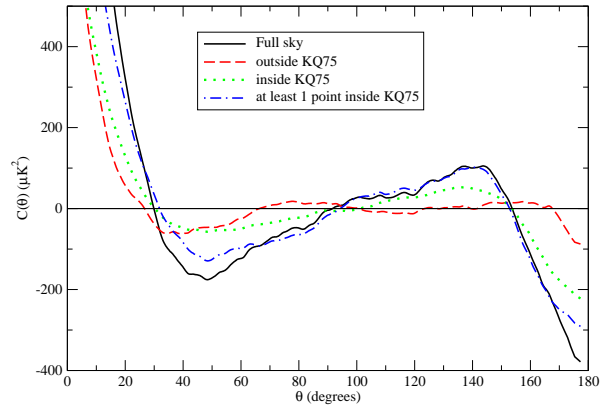}
  \caption{The two-point angular correlation function from the WMAP 5 year
    results.  Plotted are $\mathcal{C} (\theta)$ for the ILC calculated
    separately on the part of the sky outside the KQ75 cut (dashed line), inside
    the KQ75 cut (dotted line), and on the part of the sky with at least on
    point inside the KQ75 cut (dotted-dashed line).  For better comparison to
    the full-sky $\mathcal{C} (\theta)$ (solid line), the partial-sky
    $\mathcal{C} (\theta)$ have been scaled by the fraction of the sky over
    which they are calculated. Adopted from  Ref.~\cite{wmap12345}. }
  \label{fig:ctheta_contributions}
\end{figure}

The little large-angle correlation that does appear in the full sky maps (for
example the solid, black line in Fig.~\ref{fig:ctheta}) is associated with
points inside the masked region.  Shown in Fig.~\ref{fig:ctheta_contributions}
are the normalized contributions to $\mathcal{C}(\theta)$ from different parts
of the map.  In particular, we see that almost all of the contribution to the
full sky two-point angular correlation function comes from correlations with
at least one point inside the masked region.  Conversely, there is essentially
no large-angle correlation for points outside the masked region and even very
little among the points completely inside the mask.  We also see that all the
curves cross zero at nearly the same angle, $\theta\sim90^\circ$.  We have no
explanation for these results though they may point to systematics in the
data.

\subsection{Alternative Statistics}\label{sec:alt}

The two-point angular correlation function, $\mathcal C(\theta)$, as defined
above in Eq.~(\ref{eqn:Ctheta}) is a simple pixel based measure of
correlations.  It makes no assumptions about the underlying theory, which can
be taken as a feature or as a flaw.  On the positive side,
Eq.~(\ref{eqn:Ctheta}) does not assume that the standard model is correct and
try to ``force'' it on the data.  On the negative side we are not utilizing
the full information available when comparing to the standard model.  

Various approaches have been taken to incorporate the standard model in the
analysis.  For example, Hajian~\cite{Hajian2007} defined a statistic
that explicitly takes into account the covariance in the quantity $C(\theta)$:
\begin{equation}
  A(x)\equiv \int_{-1}^x\int_{-1}^x
    C(\theta)F^{-1}(\theta,\theta')C(\theta') d(\cos\theta)d(\cos\theta'),
\end{equation}
where $F$ is the aforementioned covariance
\begin{equation}
  F(\theta,\theta')\equiv \langle \left[C(\theta)-\langle
  C(\theta)\rangle\right]\left[ C(\theta')-\langle
  C(\theta')\rangle\right]\rangle,
\end{equation}
and as usual the angle brackets denote an ensemble average.  Note that in the
limit that $C(\theta)$ is uncorrelated then $A(1/2)=S_{1/2}$.  Clearly this
statistic relies on a model to calculate $F(\theta,\theta')$.  With this
statistic and assuming the concordance model it is found that less than $1\%$
of realizations of the standard model have a $A(0.53)$ less than those found
for the masked skies.  For the full sky ILC map approximately $8\%$ of
realizations have a smaller value.  Though less constraining than making no
assumptions about the theory through the use of the $S_{1/2}$ statistic, these
results are consistent with those we previously found.

Another approach advocated by Efstathiou \etal~\cite{Efstathiou2010} is to
reconstruct the full-sky $C(\theta)$ from the partial sky and compare the
reconstructed full-sky $C(\theta)$ (using, say $S_{1/2}$) to the predictions
of the model.  This approach employs the usual map making algorithm on the
low-$\ell$ spherical harmonic coefficients, $\alm$.  In this approach it is
assumed that the statistical properties of the $\alm$ above some
$\ell_{\mathrm{max}}$ are known.  In particular it assumes there are no
correlations between the $\alm$ with $\ell <\ell_{\mathrm{max}}$ and those
with $\ell >\ell_{\mathrm{max}}$.  The method is very similar to a maximum
likelihood analysis (see~\cite{Efstathiou2003}, for example).  As we have seen
above (e.g.\ in Fig.~\ref{fig:ctheta}) it is not surprising that this approach
will be consistent with the full-sky ILC results, as these authors have
verified.  As mentioned in the previous subsection, however, that this
procedure poses a {\it different} question of the data than has been addressed
by the $S_{1/2}$ statistic applied to a masked sky.  With the map-making
technique a full-sky map is constructed that is consistent with the sky
outside the mask but relies on assumptions to fill in the masked region.  As
is clear from Fig.~\ref{fig:ctheta_contributions} it is precisely the region
inside the mask that is introducing correlations with the region outside the
mask.  Thus, the assumptions required to allow filling in the masked region
also produce two-point angular correlations.

Whether or not reconstructing the full sky is a ``more optimal'' approach than
direct calculation of the cut-sky $C(\theta)$ is moot, but likely depends on
the actual (rather than the assumed) statistical properties of the underlying
fluctuations, as well as on the particular realization of those distributions.
What \textit{is} important is to make an ``apples to apples'' comparison
between the observed sky and simulated realizations of the ensemble of
possible skies.  As we have shown, the pixel-based two-point correlation
function on the region of the sky outside a conservative galactic mask is
inconsistent with the predictions of the standard $\Lambda$CDM model for the
identical pixel-based two-point correlation function on the identically masked
sky.  The fact that
  the full-sky analysis shows less statistical significance is not in
  contradiction with the cut-sky result, although it may eventually help in
  pointing to a cause of this anomaly.

\subsection{Summary}

The striking feature of the two-point angular correlation function as seen in
Fig.~\ref{fig:ctheta} is not that it disagrees with $\Lambda$CDM (though it
does at $> 90$\% C.L.)  but that at large angles it is nearly zero.  This lack
of large angle correlations is unexpected in inflationary models.  The
$S_{1/2}$ statistic quantifies the deviation from zero and shows a discrepancy
exists at more than $99.9\%$ C.L.  Equally striking is the fact that the
little correlation that does exist in the full sky ILC map, or equally in the
MLE estimated $\alm$, comes from correlations between the masked foreground
region and the expectedly cleaner CMB regions of the sky.  Thus this residual
correlation, which still is discrepant with generic inflationary predictions
at about $95\%$ C.L., comes from the reconstruction procedure.  This
surprising lack of large angle correlation outside the masked region remains
an open problem.

We also note that the vanishing of power is much more apparent in
real space (as in $\mathcal{C}(\theta)$) than in multipole space (as in
$C_\ell$).  The harmonic-space quadrupole and octopole are only moderately
low (e.g.~\cite{O'Dwyer2004}), and it is really a range of low multipoles
that conspire to make up the vanishing $\mathcal{C}(\theta)$.   Specifically,
as discussed in \cite{wmap12345}, there is a cancellation between 
the combined contributions of $C_2$,...,$C_5$ and the contributions of $C_\ell$
with $\ell\geq 6$.  It is this conspiracy that is most disturbing, since it violates
the independence of the $C_\ell$ of different $\ell$ that defines statistical isotropy.

In \cite{wmap12345} we therefore explored the possibility that the vanishing
of $C(\theta)$ could be explained simply by changing the values of the
theoretical low-$\ell$ $C_\ell$, as might be the result, say, of a modified
inflaton potential.  In particular, we replaced $C_2$ through $C_{20}$ in the
best-fit $\Lambda$CDM model with the values extracted from the cut-sky ILC
five-year map. From these $C_\ell$, $200,000$ random maps were created,
masked, and $S_{1/2}$ computed. Under the assumptions of Gaussianity and
statistical isotropy of these $C_\ell$ only 3 per cent of the generated maps
had $S_{1/2}$ less than the cut-sky ILC5 value. Thus, even if the $C_\ell$ are
set to the specific values that produce such a low $S_{1/2}$, a Gaussian
random, statistically isotropic realization is unlikely to produce the
observed lack of large angle correlations at the 97\% C.L.  Moreover, in work
in progress we show that almost all of those $3\%$ are skies with several
anomalously low $C_\ell$ --- not at all the sky we see.  Only a tiny fraction
of the $3\%$ represent skies in which most of the individual $C_\ell$ were
close to the $\Lambda$CDM prediction but they conspire to cancel one another
in the large angle $C(\theta)$.  This shows that either (i) the low-$\ell$
$C_\ell$ are correlated, contrary to the assumption of statistical
isotropy\footnote{Though this appears to be an unlikely explanation since
  correlations between the $C_\ell$ generically increase the variance of the $S_{1/2}$
  statistic \cite{Pontzen_Peiris}.}  or (ii) our Universe is an extremely
unlikely realization of whatever statistically isotropic model one devises.

It is for this reason that theoretical efforts to explain ``low power on large
scales'' must focus on explaining the low $\mathcal{C}(\theta)$ at
$\theta\gtrsim 60$ deg, rather than the low quadrupole.

Finally, one might ask if the observed lack of correlation and the alignment
of quadrupole and octopole are correlated. This issue was studied by Raki\'c
\& Schwarz \cite{Rakic2007} for the full sky and by Sarkar \etal~\cite{Sarkar}
for the cut sky case. In both cases, it was shown that low power and
alignments are uncorrelated --- i.e.\ that having one does not imply a larger
or smaller probability of having the other.  
This was shown by applying a Monte-Carlo analysis to sky realizations with the
underlying standard Gaussian random, statistically isotropic cosmological
model, without any further constraints.  Thus one might view the $99.6$\%
C.L. of quadrupole-octopole alignment presented in the previous section and
the $95$\% C.L. for lack of correlation in full-sky maps reported in this
section as statistically independent.

\section{Quest for an explanation}
\label{sec:explain}

Understanding the origin of CMB anomalies is clearly important.  
Both the observed alignments of the low-$\ell$ full-sky multipoles,
and the absence of large-angle correlations (especially on the galaxy-cut sky) 
are severely inconsistent with predictions of standard cosmological theory. 
There are four classes of possible explanations:
\begin{enumerate}
\item Astrophysical foregrounds,
\item Artifacts of faulty data analysis,
\item Instrumental systematics,
\item Theoretical/cosmological. 
\end{enumerate}

In this section, we review these four classes of explanations, giving
examples from each. First, however, we discuss two generic ways to break
statistical isotropy and affect the intrinsic (true) CMB signal --- additive
and multiplicative modulations --- and illustrate in general terms why it has
been so difficult to \textit{explain} the anomalies.

\subsection{Additive vs.\ multiplicative effects}
\label{sec:general}

Why is it difficult to explaining the observed CMB anomalies? There are three
basic reasons:

\begin{itemize}
\item Most explanations work by \textit{adding} power to the large-angle CMB, while the
  observed anisotropies actually have \textit{less} large-scale power, and particularly less
  large angle correlation, than the
  $\Lambda$CDM cosmological model predicts.
 \item Unaccounted for sources of CMB fluctuations in the foreground, even if
   possessing/causing aligned low-$\ell$ multipoles of their own, cannot bring
   unaligned statistically isotropic cosmological perturbations into
   alignment.  Therefore, aligned foregrounds as an explanation for alignment
   work only if the cosmological signal is subdominant, thus exacerbating the
   lack of large-angle correlations.
\item The alignments of the quadrupole and octopole are with respect to the
  ecliptic plane and near the dipole direction. It is generally difficult
  to have these directions naturally be picked out by any class of
  explanations (though there are exceptions to this --- see the
  instrumental example below).
\end{itemize}

Gordon \etal~\cite{Gordon2005}, Raki\'c \& Schwarz \cite{Rakic2007} and Bunn
\& Bourdon \cite{Bunn_Bourdon} explored generic ``additive models'' where the
temperature modification that causes the alignment is added to the intrinsic
temperature
\begin{equation}
T_{\mathrm{observed}}(\hat e)=T_{\mathrm{intrinsic}}(\hat
e)+T_{\mathrm{add}}(\hat e). 
\label{eq:add}
\end{equation}
Here $T_{\mathrm{add}}(\hat e)$ is the additive term --- perhaps contamination
by a foreground, perhaps an additive instrumental or cosmological effect. They
showed that additive modulations of the CMB sky that ameliorate the alignment
problems tend to \textit{worsen} the overall likelihood at large scales
(though they may pick up offsetting positive likelihood contribution from
higher multipoles).  The intuitive reason for this is that there are two
penalties incurred by the additive modulation. First, since the power spectrum
at low $\ell$ is lower than expected, one typically needs to arrange for an
accidental cancellation between $T_{\mathrm{intrinsic}}$ and
$T_{\mathrm{add}}$; the cancellation moreover must leave an aligned quadrupole
and octopole even though the quadrupole and octopole of
$T_{\mathrm{intrinsic}}$ are not aligned.  (Very similar reasoning argues
against additive explanations of the suppression of large angle correlations.)
Second, the simplest models for the additive contribution that are based on an
azimuthally symmetric modulation of a gradient field can only affect $m=0$
multipoles around the preferred axis, while as we mentioned earlier the
observed quadrupole and octopole as seen in the preferred (dipole) frame are
dominated by the $m=\ell$ components.

\begin{figure}[!t]
\includegraphics[scale=0.30]{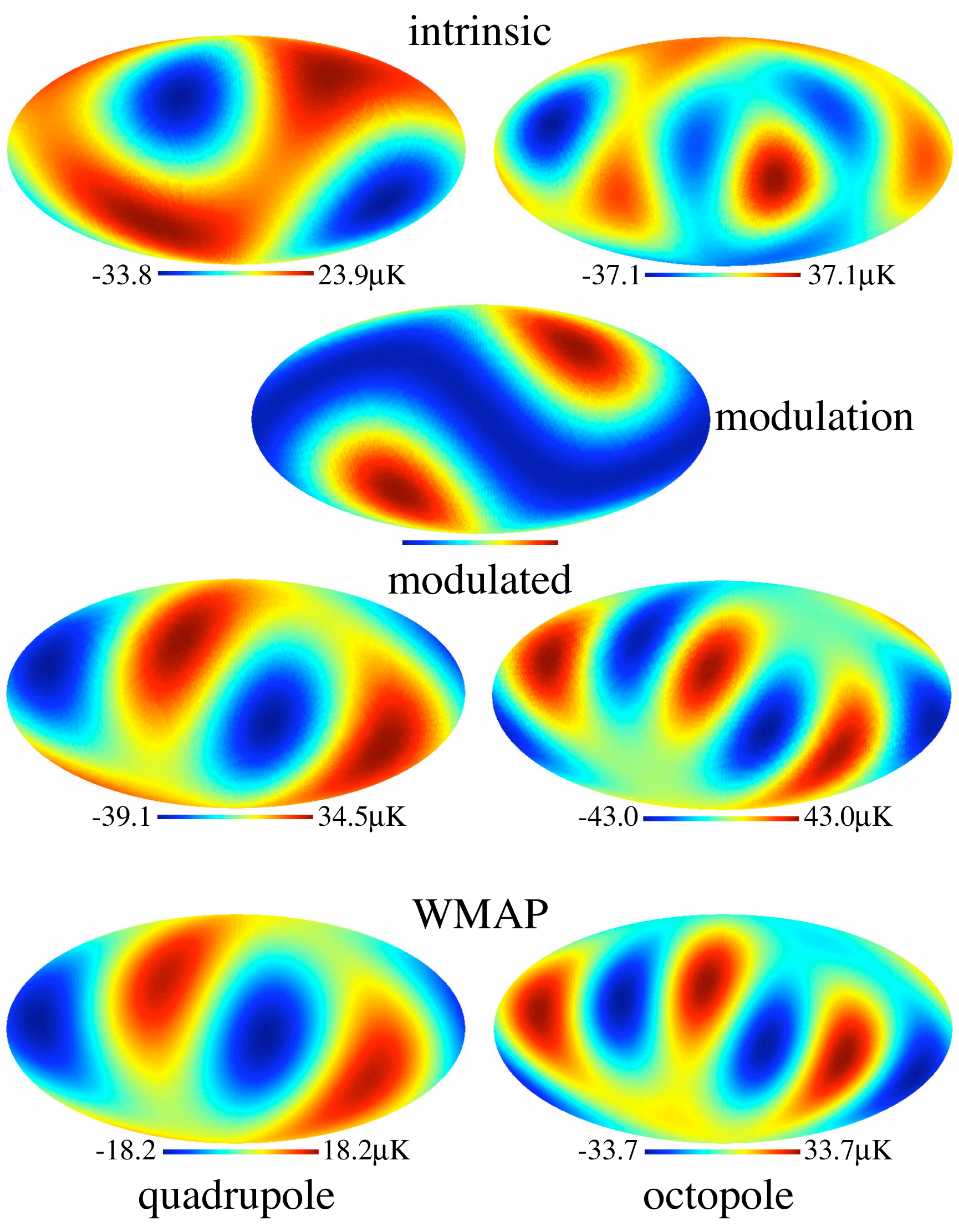}
\caption{A realization of the multiplicative model where the quadrupole (left
  column) and octopole (right column) exhibit an alignment similar to WMAP.
  First row: intrinsic (unmodulated) sky from a Gaussian random isotropic
  realization.  Second row (single column): the quadrupolar modulation with
  $f=-1$ and $w_2=-7$ (see Eq.~(\ref{eq:mult})) in the dipole direction.
  Third row: the modulated sky of the observed CMB.  Fourth row: WMAP full-sky
  quadrupole and octopole. Adopted from Ref.~\cite{Gordon2005}.  }
\label{fig:mult}
\end{figure} 

In contrast to the additive models, the multiplicative mechanisms, where the
intrinsic temperature is multiplied by a spatially varying modulation, are
phenomenologically more promising. As a proof of principle, a toy-model
modulation \cite{Gordon2005}
\begin{equation}
T_{\mathrm{observed}}(\hat e)=f\left [1+w_2Y_{20}(\hat e)\right ]\,
T_{\mathrm{intrinsic}}(\hat e),
\label{eq:mult}
\end{equation}
(where the modulation is a pure $Y_{20}$ along the dipole axis), can increase
the likelihood of the WMAP data by a factor of $\exp{(16/2)}$ and, at the same
time, increase the probability of obtaining a sky with more alignment
(e.g.\ higher angular momentum statistic) 200 times, to 45\%; see
Fig.~\ref{fig:mult}. Indeed, Groeneboom \etal~\cite{Groeneboom:2009qf},
building on the work of Groeneboom \& Eriksen \cite{Groeneboom:2008fz} and
Hanson \& Lewis \cite{Hanson:2009gu} and motivated by a model due to Ackerman
\etal~\cite{Ackerman:2007nb}, have identified a $9\sigma$ quadrupolar power
asymmetry, recently confirmed by the WMAP team \cite{Bennett:2010jb}; this
anomaly can, however, be fully explained by accounting for asymmetric beams
\cite{Hanson_beams}.  Recently, Hoftuft \etal ~\cite{Hoftuft} found a greater
than 3-$\sigma$ evidence for nonzero dipolar modulation of the power.

\subsection{Astrophysical explanations}
\label{sec:foregr}

One fairly obvious possibility is that there is a pernicious foreground that
contaminates the primordial CMB and leads to the observed anomalies.  Such
foregrounds are, of course, additive mechanisms, in the sense of the preceding
section, and so suffer from the shortcomings described therein.  Moreover,
most such foregrounds are Galactic, while the observed alignments are
with respect to the ecliptic plane. One would expect that Galactic foregrounds
should lead to Galactic and not ecliptic foregrounds. This simple expectation
was confirmed in \cite{lowl2}, where we showed that, by artificially adding a
large admixture of Galactic foregrounds to WMAP CMB maps, the quadrupole
vectors move near the $z$-axis and the normal into the Galactic plane, while
for the octopole all three normals become close to the Galactic disk at
$90^\circ$ from the Galactic center. Therefore, as expected Galactic
foregrounds lead to Galactic, and not ecliptic, correlations of the quadrupole
and octopole (see also studies by \cite{SS2004,Bielewicz2005}).

Moreover, in \cite{lowl2}, we have shown that the known Galactic foregrounds
possess a multipole vector structure very different from that of the observed
quadrupole and octopole. The quadrupole is nearly pure $Y_{22}$ in
the frame where the $z$-axis is parallel to the dipole (or $\hat w^{(2,1,2)}$
or any nearly equivalent direction), while the octopole is dominantly $Y_{33}$
in the same frame.  Mechanisms which produce an alteration of the microwave
signal from a relatively small patch of sky --- and all of the recent proposals
fall into this class --- are most likely to produce aligned $Y_{20}$ and
$Y_{30}$. This is essentially because the low-$\ell$ multipole vectors 
will all be parallel to each other, leading to a $Y_{\ell 0}$ in this frame.

A number of authors have attempted to explain the observed quadrupole-octopole
correlations in terms of a \textit{new} foreground --- for example the
Rees-Sciama effect \cite{Rakic2006,Rakic2007}, interstellar dust
\cite{Frisch05}, local voids \cite{Inoue_Silk}, or the Sunyaev-Zeldovich
effect \cite{Peiris_Smith}. Most if not all of these proposals have a
difficult time explaining the anomalies without severe fine tuning.  For
example, Vale \cite{Vale05} cleverly suggested that the moving lens effect,
with the Great Attractor as a source, might be responsible for the extra
anisotropy; however Cooray \& Seto \cite{Cooray05} have argued that the
lensing effect is far too small and requires too large a mass of the
Attractor. 

It is also interesting to ask if any known or unknown Solar system physics
could lead to the observed alignments. Dikarev \etal~
\cite{Dikarev2008,Dikarev2009} studied the question of whether solar system
dust could give rise to sizable levels of microwave emission or
absorption. Surprisingly, very little is known about dust grains of mm to cm
size in the Solar system, and their absorption/emission properties strongly
depend on their chemical composition. While iron and ice particles can
definitely be excluded to contribute at significant levels, carboneous and
silicate dust grains might contribute up to a few $\mu$K close to the ecliptic
plane, e.g.~due to the trans-Neptunian object belt.  Such an extra
contribution along the ecliptic could give rise to CMB structures aligned with
the ecliptic, but those would look very different from the observed ones. On
top of that, Solar system dust would be a new additive foreground and could
not explain the lack of large angle correlations.  Thus it seems unlikely that
Solar system dust grains cause the reported large angle anomalies,
nevertheless they are sources of microwave absorption and emission and may
become important to precision measurements in the future.

Finally, it has often been suggested to some of us in private communications
that the anomalies may not reflect an unknown foreground that has been
neglected, but rather the ``mis-subtraction'' of a known foreground. However,
it has never quite been clear to us how this leads to the observed alignments
or lack or large angle correlations, and we are unaware of any literature that
realizes this suggestion successfully.

\subsection{Data analysis explanations}
\label{sec:analysis}

Most of the results discussed so far have been obtained using reconstructed
full-sky maps of the WMAP observations \cite{Bennett2003,TOH,LILC}.  In the
presence of the sky cut of even just a few degrees, the errors in the
reconstructed anisotropy pattern, and the directions of multipole vectors, are
too large to allow drawing quantitative conclusions about the observed
alignments \cite{Copi2004}. These large errors are expected: while the
\textit{power} in the CMB (represented, say, by the angular power spectrum
$C_\ell$) can be accurately recovered since there are $2\ell+1$ modes
available for each $\ell$, there are only 2 modes available for each multipole
vector; hence the cut-sky reconstruction is noisier. However the cut-sky
alignment probabilities, while very uncertain, are consistent with the
full-sky values \cite{lowl2,Bielewicz2005}; more generally, the alignments
appear to be rather robust to Galactic cuts and foreground contamination
\cite{deOliveiraCosta:2006zj}.

A different kind of explanation of missing large-scale power, or missing large
angle correlations, has been taken by Efstathiou \etal~\cite{Efstathiou2010}
who argued that maximum likelihood estimators can be applied to the cut-sky
maps to reliably and optimally reconstruct the CMB anisotropy of the whole
sky; for a recent work that extends these ideas, see
\cite{Pontzen_Peiris}. This approach yields more two-point correlation on
large scales ($S_{1/2}\sim 8000\,(\mu K)^4$, which is $\sim 5\%$ likely) than
the direct cut-sky pixel-based calculation which gives $S_{1/2}\sim 1000\,(\mu
K)^4$ result and is $\sim 0.03\%$ likely. These authors then argue that the
extremely low $S_{1/2}$ obtained by the pixel-based approach applied to the
cut sky is essentially a fluke, and the more reliable result comes from their
maximum-likelihood reconstruction of the full sky. It may indeed be true that
the pixel-based calculation is a suboptimal estimate of the {\it full-sky}
$C(\theta)$ \textit{for a statistically isotropic cosmology}.  However, quantities
calculated on the cut sky are clearly
insensitive to assumptions about what lies behind the cut. We can only observe
reliably the $\sim 75\%$ of the sky that was not masked, and that is where the
large-angle two-point-correlation is near-vanishing. Any attempt to
reconstruct the full sky must make assumptions about the statistical
properties of the CMB sky, and would clearly be affected by the coupling of
small-scale and large-scale modes --- exactly what is necessary to have a sky
in which $S_{1/2}$ is anomalously low, while the $C_\ell$s are individually
approximately consistent with the standard cosmology.

\subsection{Instrumental explanations}
\label{sec:instr}

Are instrumental artifacts the cause of the observed alignments (and/or the
low large-scale power)? One possible scenario would go as follows.  WMAP
avoids making observations near the Sun, therefore covering regions away from
the ecliptic more than those near the ecliptic. While the corresponding
variations in the noise per pixel are well known (e.g.\ as the number of
observations per pixel, $N_{\mathrm{obs}}$; see Fig.~3 in \cite{Bennett2003}), and
its effects on the large-scale anomalies are ostensibly small, they could, in
principle, be amplified and create the observed ecliptic anomalies.  However a
successful proposal for such an amplification has not yet been put forward.

Another possibility is that an imperfect instrument couples with dominant
signals from the sky to create anomalies. Let us review an example given in
Ref.~\cite{Gordon2005}: suppose that the instrumental response
$T_{\mathrm{instr}}(\hat e)$ to the true sky signal $T(\hat e)$ is nonlinear
\begin{equation}
T_{\mathrm{instr}}(\hat e) 
=f  \sum_i \alpha_i  \left[ {T(\hat e) \over f }\right] ^i  \,.
\end{equation}
Here $f$ is an arbitrary normalization scale for the non-linearity of the
response, and $\alpha_i$ are arbitrary coefficients with $\alpha_1=1$.  If
$\alpha_{i>1} \ne 0$ then $T_{\mathrm{instr}} \ne T$ and the observed temperature
is a non-linear modulation of the true temperature. The dominant temperature
signal for a differencing experiment such as WMAP is the dipole arising from
our peculiar motion, $T(\hat e)=T_{\mathrm{di}p}\cos\theta$, with $T_{\mathrm
    {dip}}=3.35$mK and $\theta$ the polar angle in the dipole frame.  Taking
$f=T_{\mathrm{dip}}$,
\begin{eqnarray}
{T_{\mathrm{instr}}(\hat e)\over T_{\mathrm{dip}}} &=& 
\alpha_1 P_1(\cos\theta) + \alpha_2 \left [{2 \over 3}P_2(\cos\theta) +1\right ] 
   \\ 
&&  + \alpha_3 \left [{2\over 5} P_3(\cos\theta) + {3 \over 5} P_1(\cos\theta)\right ] 
+ \ldots \,, \nonumber
\end{eqnarray}
where $P_\ell$ are the Legendre polynomials.  Note that with $\alpha_2 \sim
\alpha_3 \sim O(10^{-2})$, the $10^{-3}$ dipole anisotropy is modulated into a
$10^{-5}$ quadrupole and octopole anisotropy which are aligned \textit{in the
  dipole frame} with the $m=0$ multipole structure.  Unfortunately (or
fortunately!), WMAP detectors are known to be linear to much better than 1\%,
so this particular realization of the instrumental explanation does not
work. As an aside, note that this type of explanation needs to assume that the
higher multipoles are not aligned with the dipole/ecliptic, and moreover,
requires essentially no intrinsic power at large scales (that is, even less
than what is observed).

To summarize: even though the ecliptic alignments (and the north-south power
asymmetry) hint at a systematic effect due to some kind of coupling of an
observational strategy and the instrument, to date no plausible proposal of this
sort has been put forth.

\subsection{Cosmological Explanations}

The most exciting possibility is that the observed anomalies have primordial
origin, and potentially inform us about the conditions in the early universe. 
One expects that in this case the alignments with the dipole, or with the solar system,
would be statistical flukes.

The breaking of statistical isotropy implies that the usual relation $\langle
\alm^* \alm\rangle=C_\ell \delta_{\ell \ell'}\delta_{mm'}$ does not hold any
more; instead
\begin{equation}
\langle \alm^* \alm\rangle=C_{\ell\ell' mm'},
\end{equation}
where the detailed form of the quantity on the right-hand side is
model-dependent (see the more detailed discussions in
\cite{Hajian2003,Hajian2004,Pereira_Abramo_09}).

There are many possibilities for how the absence of statistical isotropy might
arise.  For example, in a non-trivial spatial topology, the fundamental domain
would not be rotationally invariant, and so the spherical harmonics (times an
appropriate radial function) would not be a basis of independent eigenmodes of
the fluctuations.  This would certainly lead to a correlation among $a_{\ell
  m}$ of different $m$ and also $\ell$, although not necessarily to aligned
multipoles.  The hope would be that the shape of the fundamental domain would
lead to these alignments, while a lower density of states at long wavelength
(compared to the covering space) would lead to the absence of large angle
correlations.  No specific model has been suggested to accomplish all of
those, and the matter is complicated by known bounds on cosmic topology
\cite{Cornish:1997rp,Cornish:1997hz, Cornish:1997ab,
  Cornish:2003db,ShapiroKey:2006hm} which force the fundamental domain to be
relatively large.

Alternately, in the early universe, asymmetry in the stress-energy tensor of
dark energy \cite{Battye09}, or a long-wavelength dark energy density mode
with a gradient in the desired direction \cite{Gordon2005}, could both imprint
the observed alignments via the integrated Sachs-Wolfe mechanism; but it is
hard to see how they would explain the lack of large angle correlations.  The
authors of \cite{Afshordi:2008rd} have put forward a model where the
Sachs-Wolfe contribution to low-$\ell$ multipoles is partly cancelled by the
Integrated Sachs-Wolfe contribution, but which still fails to explain the
lowness of $S_{1/2}$ or the alignments of low-$\ell$ multipoles. While models
where the anomalies are caused by breaking of the statistical isotropy
\cite{Jaffe2005,Ghosh} are especially well studied, see \cite{Carroll_transl}
for the equally interesting possibility that the \textit{translational}
invariance is broken.
  
A commonly used mechanism to explain such anomalies are inflationary models
that contain implicit breaking of isotropy
\cite{Gordon2005,ACW,ArmendarizPicon07,Gumrukcuoglu,Rodrigues_Bianchi,
  Pitrou08,Erickcek_hem,Erickcek_iso}.  Pontzen~\cite{Pontzen} helpfully shows
temperature and polarization patterns caused by various classes of Bianchi
models that explicitly break the statistical isotropy. However, outside of
explaining the anomalies, the motivation for these anisotropic models is not
compelling and they seem somewhat contrived. Moreover, the authors have not
investigated whether the low $S_{1/2}$ is also observed.  Nevertheless, given
the large-scale CMB observations, as well as the lack of fundamental theory
that would explain inflation, investigating such models is well worthwhile.

A very reasonable approach is to describe breaking of the isotropy with a
phenomenological model, measure the parameters of the model, and then try to
draw inferences about the underlying physical mechanism. For example, a
convenient approach is to describe the breaking of isotropy via the
direction-dependent power spectrum of dark matter perturbations
\cite{Pullen_Kam}
\begin{equation}   
     P(\vec{k})= A(k) 
     \left[1+ \sum_{\ell m} g_{\ell m}(k) Y_{\ell m}(\hat{k}) \right],
\end{equation}
where $k=|\vec k|$, $g_{\ell m}(k)$ quantifies the departure from statistical
isotropy as a function of wavenumber $k$, and $A(k)$ refers to the
statistically isotropic part of the power spectrum. In this model, the power
spectrum, normally considered to depend only on scale $k$, now depends on
direction in a parametric way. Statistically significant finding that $g_{\ell
  m}\neq 0$ for any $(\ell, m)$ would signal a violation of statistical
isotropy.

As with the other attempts to explain the anomalies, we conclude that, while
there have been some interesting and even promising suggestions, no
cosmological explanation to date has been compelling. 

\subsection{Alignment explanations: what next}

While future WMAP data is not expected to change any of the observed results,
our understanding and analysis techniques are likely to improve.  The most
interesting test will come from the Planck satellite, whose temperature maps,
obtained with a completely different instrument and observational technique
than WMAP, could shed significant new light on the alignments. Moreover,
polarization information could be extremely useful in distinguishing between
different models and classes of explanations in general; for example, Dvorkin
\etal~\cite{Dvorkin} explicitly show how polarization information expected
from Planck can help identify the cause of the alignments.  Finally, one could
use the large-scale structure (i.e.\ galaxy distribution) data on the largest
observable scales from surveys such as Dark Energy Survey (DES) and Large
Synoptic Survey Telescope (LSST) to test cosmological explanations; see
e.g.~\cite{Zhan2005,Pullen_Hirata}.

\section{Explanations from the WMAP team}
\label{sec:discussion}

In their seven year data release the WMAP team explicitly discusses several
CMB anomalies \cite{Bennett:2010jb} including the two main ones described
in this review.  For the first major issue --- the alignment of low
multipoles with each other --- the WMAP team agrees that the alignment is
observed and argue, based on work by Francis and Peacock
\cite{Peacock2009}, that the integrated Sachs-Wolfe (ISW) contribution of
structures at small redshifts ($z \ll 1$) could be held responsible.  There
are serious problems with this argument.  Firstly, 
the ordinary Sachs-Wolfe (SW) effect typically dominates at these $\ell$ over
the ISW.  Thus, only if the ordinary SW effect on the last scattering surface
is anomalously low will the ISW contribution dominate. Secondly, though the
ISW may lead to alignment of the quadrupole and octopole it is not an
explanation for the observed Solar system alignments.  This alignment would
need to be an additional statistical fluke.  Finally, this explanation does
nothing whatsoever to mitigate the lack of large scale angular correlation
because the ISW effect acts as an additive component and should be
statistically uncorrelated from the primordial CMB. Therefore, even if the ISW
reconstruction is taken as reliable, this argument would imply;
\begin{enumerate}
\item an accidental downward fluctuation of the SW sufficient for the ISW
  of local structure to dominate and cause an alignment, and
\item an accidental cancellation in angular correlation between the SW and ISW
  temperature patterns.
\end{enumerate}
Neither the WMAP team nor Francis and Peacock estimate the likelihood of
these two newly created puzzles.

Regarding the second major issue --- the lack of angular correlation --- the
WMAP team refers to a recent work by Efstathiou, Ma and Hanson
\cite{Efstathiou2010} who argue that quadratic estimators are better estimates
of the full sky from cut-sky data and are in better agreement with the
concordance model. While these estimators have been shown to be optimal under
the assumption of statistical isotropy, it is unclear why they should be
employed when this assumption is to be tested. (For a contrary view, see
\cite{Pontzen_Peiris}.)  The pixel based estimator applied to the cut sky
in our analysis does not rely on statistical isotropy and it is more
conservative as it does not try to reconstruct temperature anisotropies inside
the cut. Finally, our claims that the pixel based cut-sky two-point
correlation function is highly anomalous rely on comparisons to the identical
correlation function calculated on simulated cut skies.  Whether or not it is
a good estimate of the full-sky correlation function is answering a different
question, (see Sec.~\ref{sec:missing}).  Conversely, the estimator suggested
in \cite{Efstathiou2010} assumes that whatever is within the cut can be
reconstructed reliably by truncating the number of multipole moments
considered. The latter logic is equivalent to the assumption of the
statistical independence of low and high multipoles, which is exactly a
consequence of statistical isotropy.

These arguments from the WMAP team offer neither new nor convincing
explanations of the observed anomalies discussed in this review.  At best
they replace one set of anomalies for another.
 
\section{Conclusions}

The CMB is
widely regarded as offering strong substantiating evidence for the concordance
model of cosmology.  Indeed the agreement between theory and data is
remarkable --- the patterns in the two point correlation functions (TT, TE and
EE) of Doppler peaks and troughs are reproduced in detail by fitting with only
six (or so) cosmological parameters.  This agreement should not be taken
lightly; it shows our precise understanding of the causal physics on the last
scattering surface. Even so, the cosmological model we arrive at is baroque,
requiring the introduction at different scales and epochs of three sources of
energy density that are only detected gravitationally --- dark matter, dark
energy and the inflaton.  This alone should encourage us to continuously
challenge the model and probe the observations particularly on scales larger
than the horizon at the time of last scattering.

At the very least, probes of the large-angle (low-$\ell$) properties of the
CMB reveal that we do not live in a typical realization of the concordance
model of inflationary $\Lambda$CDM.  We have reviewed a number of the ways in
which that is true: the peculiar geometry of the $\ell=2$ and $3$ multipoles
--- their planarity, their mutual alignment, their alignment perpendicular to
the ecliptic and to the dipole; the north-south asymmetry; and the near
absence of two-point correlations for points separated by more than $60^o$.

If indeed the observed $\ell=2$ and $3$ CMB fluctuations are not cosmological,
one must reconsider all CMB results that rely on the low $\ell$, e.g.~the
measurement of the optical depth from CMB polarization at low $\ell$ or the
spectral index of scalar perturbations and its running.  Moreover, the
CMB-galaxy cross-correlation, which has been used to provide evidence for the
Integrated Sachs-Wolfe effect and hence the existence of dark energy, also
gets contributions from the lowest multipoles (though the main contribution
comes from slightly smaller scales, $\ell\sim 10$).  Indeed, it is quite
possible that the underlying physical mechanism does not cut off abruptly at
the octopole, but rather affects the higher multipoles.  Indeed, several
pieces of evidence have been presented for anomalies at $l>3$ (e.g.\
\cite{Land2005a,Land:2006bn}). One of these is the parity of the microwave
sky.  While the observational fact that the octopole is larger than the
quadrupole ($C_3 > C_2$) is not remarkable on its own, including higher
multipoles (up to $\ell \sim 20$) the microwave sky appears to be parity odd
at a statistically significant level (since WMAP 5yr)
\cite{Land:2005jq,Kim:2010gf,Kim:2010gd}. It is hard to imagine a cosmological
explanation for a parity odd universe, but the same holds true for
unidentified systematics or unaccounted astrophysical foregrounds, especially
as this recently noticed puzzle shows up in the very well studied angular
power spectrum.

While the further WMAP data is not expected to change any of the observed
results, our understanding and analysis techniques are likely to improve. Much
work remains to study the large-scale correlations using improved foreground
treatment, accounting even for the subtle systematics, and in particular
studying the time-ordered data from the spacecraft. The Planck experiment will
be of great importance, as it will provide maps of the largest scales obtained
using a very different experimental approach than WMAP --- measuring the
absolute temperature rather than temperature differences. Polarization maps,
when available at high enough signal-to-noise at large scales (which may not
be soon), will be a fantastic independent test of the alignments, as each
explanation for the alignments, in principle, also predicts the statistics of
the polarization pattern on the sky.

\acknowledgments

DH is supported by DOE OJI grant under contract DE-FG02-95ER40899, and NSF
under contract AST-0807564. DH and CJC are supported by NASA under contract
NNX09AC89G; DJS is supported by Deutsche Forschungsgemeinschaft (DFG); GDS is
supported by a grant from the US Department of Energy; both GDS and CJC are
supported by NASA under cooperative agreement NNX07AG89G.

\bibliography{cmb_review}

\end{document}